\documentstyle[11pt,twoside,jltp,epsfig]{article}

\runninghead{F. Pierre, H. Pothier, D. Esteve, and M.H. Devoret}{Energy Redistribution Between Quasiparticles in Silver Wires}
\input tcilatex
\begin{document}

\title{Energy Redistribution Between Quasiparticles in Mesoscopic Silver Wires}
\author{F.\ Pierre%
\address{ Service de Physique de l'Etat Condens\'{e}, Commissariat
\`{a} l'Energie Atomique, Saclay, F-91191 Gif-sur-Yvette, France}, H.
Pothier, D. Esteve, and M.H. Devoret}

\begin{abstract}
We have measured with a tunnel probe the energy distribution function of
quasiparticles in silver diffusive wires connected to two large pads
(``reservoirs''), between which a bias voltage was applied. From the
dependence in energy and bias voltage of the distribution function we have
inferred the energy exchange rate between quasiparticles. In contrast with
previously obtained results on copper and gold wires, these data on silver
wires can be well interpreted with the theory of diffusive conductors either
solely, or associated with another mechanism, possibly the coupling to
two-level systems.

\end{abstract}

\maketitle

\vspace{0.3in}

The present understanding of metals at low temperature relies on Landau's
theory of Fermi liquids. In this theory, the elementary excitations of the
electron fluid are nearly independent fermionic quasiparticles\cite{Pines}.
The residual interactions depend on the efficiency of the screening of
Coulomb interactions, and increase if electrons are scattered by impurities,
surface or lattice defects \cite{Schmid, Altshuler}. These interactions can
be probed through the shape of the energy distribution function in an
out-of-equilibrium situation. We have found in previous experiments on copper%
\cite{Relax_1} and gold\cite{gold} wires that the energy exchange rate
between quasiparticles is stronger and has a different energy dependence
than predicted by Altshuler and Aronov's (AA) theory of diffusive conductors%
\cite{Altshuler}. In the present paper, we report measurements of
distribution functions in silver wires, where the interactions appear to be
in closer agreement with AA theory.
\begin{figure}[!tb]
\centering
\epsfig{file=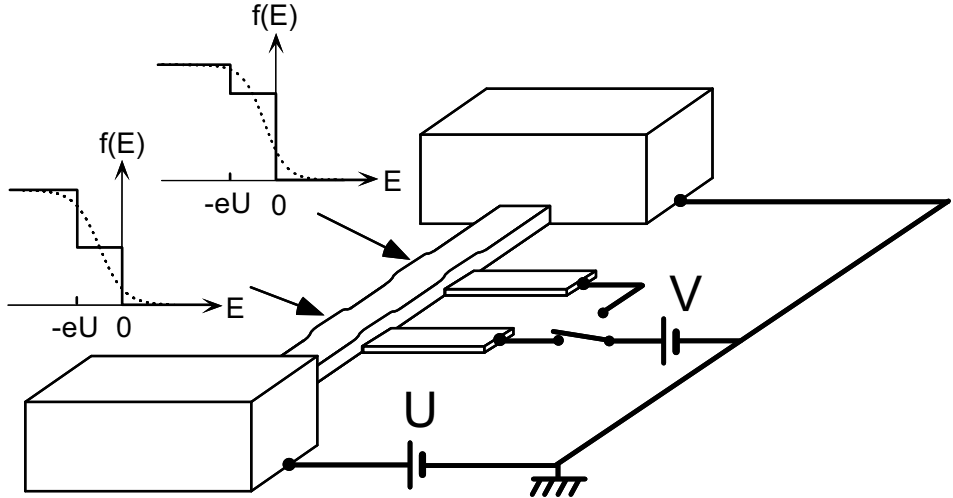, width=4.9692in, clip=}
\caption{Experimental layout: a metallic wire of length $L$ is connected to
large reservoir electrodes, biased at potentials $0$ and $U$. In absence of
interaction, the distribution function at a distance $X=xL$ from the
grounded electrode has an intermediate step $f\left( E\right) =1-x$ for
energies between $-eU$ and $0$ (solid curves) (we assume $U>0$). When
interactions are strong enough to thermalize electrons, the distribution
function is a Fermi function, with a space-dependent temperature and
electrochemical potential (dotted curves). In the experiment, the
distribution function is obtained from the differential conductance $%
dI/dV(V) $ of the tunnel junction formed by the wire and a superconducting
electrode placed underneath.}
\label{FigSetup}
\end{figure}


The experimental setup used to probe the energy exchange rate is shown in
Fig.~\ref{FigSetup}. A metallic diffusive wire of length $L$ is connected at
both ends to large and thick electrodes called ``reservoirs'' in the
following. The quasiparticle energy distribution $f(x,E)$ at a distance $%
X=xL $ from the right electrode, is obtained from the differential
conductance $dI/dV\left( V\right) $ of a tunnel junction between the wire
and a superconducting electrode\cite{Relax_1, Rowell}. A voltage difference $%
U$ is applied between the reservoirs in order to implement a stationary
out-of-equilibrium situation. The shape of the distribution function $f(x,E)$
depends on the average number of inelastic collisions a quasiparticle
experiences during its diffusive motion from one of the electrodes to the
position $X,$ and on the amount of energy exchanged at each collision.

We present the results of four experiments, labeled A, B, C and D in the
following. The wire length $L=$ $5,$ $10,$ or $20\ {\rm \mu m}$ is appended
to the label; for instance the samples A$5$ and C$20$ are $5$ and $20\ {\rm %
\mu m}$-long, respectively. All samples were fabricated by electron-gun
evaporation of silver at several angles through a PMMA suspended mask
patterned using e-beam lithography. The substrate was, as in our experiments
on Cu and Au, thermally oxidized silicon. Samples with same labels (B$5$ and
B$10$ on the one hand, D$20{\rm a}$ and D$20{\rm b}$ on the other hand) were
fabricated simultaneously. The thickness of the wires is $45~{\rm nm}$ and
their width $w$ ranges between $65$ ${\rm nm}$ and $150$ ${\rm nm}.$ The
electrodes at the ends of the wires are $400~{\rm nm}$-thick silver pads
with an area of about $1~{\rm mm}^2,$ thereby implementing adequate
reservoirs \cite{Landauer}. The superconducting probes, made of aluminum 
\cite{Aleiner}, were positioned at $x=0.5$ and/or at $x=0.2.$ The areas of
the tunnel junctions are $w\times 150\ {\rm nm}$ and their tunnel
resistances range between $23~{\rm k}\Omega $ and $130~{\rm k}\Omega .$ We
estimated the diffusion coefficient $D$ of quasiparticles, and hence the
diffusion time $\tau _D=L^2/D,$ from the low-temperature resistance of the
wires. The samples were mounted in a copper box thermally anchored to the
mixing chamber of a dilution refrigerator. Electrical connections were made
through filtered coaxial lines \cite{filtres}, and measurements were carried
out at a temperature of 40~mK.

\begin{figure}[!tb]
\centering
\epsfig{file=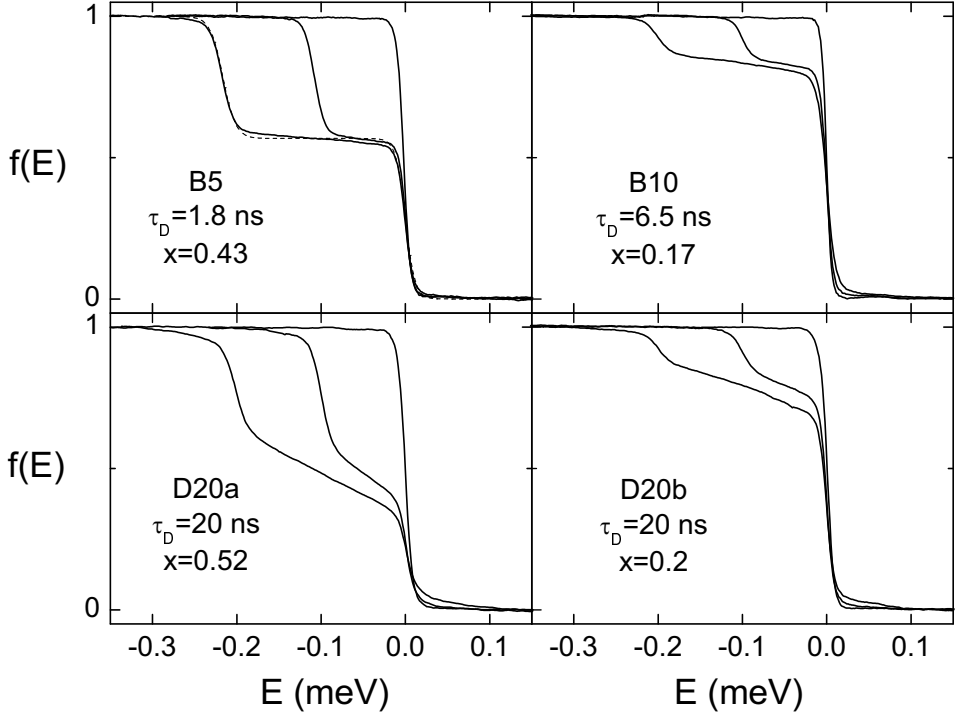, width=4.9787in, clip=}
\caption{
Measured distribution functions for $U=0,$ $0.1$ and $0.2~{\rm mV}$ in
samples B5, B10, D20a and D20b (see Table 1). In the top left panel, the
dotted line is the prediction for the non-interacting regime (Eq.~(2)) for $%
U=0.2~{\rm mV.}$}
\label{FigData}
\end{figure}


The distribution functions obtained at $U=0,$ $0.1$ and $0.2$~${\rm mV}$ are
shown in Fig.~\ref{FigData}. For $U=0,$ $f(E)$ is close to the expected
Fermi function at the temperature of the refrigerator. For $U\neq 0$, the
functions $f(E)$ measured near the middle of the wire B$5$ (top left panel)
display a sharp double step with a plateau at height $1-x=0.57$. For
comparison, we have plotted as a dotted line the best fit at $U=0.2~{\rm mV}$
with a linear combination of the Fermi functions of the reservoirs\cite
{Treservoir}, which is the expected distribution function in the
non-interacting regime (see Refs.\cite{Relax_1, Nagaev} and Eq. (\ref{fzero}%
) below). The deviations from this regime are more apparent in sample D$20%
{\rm a}$ (bottom left panel), for which the diffusion time $\tau _D=20\ {\rm %
ns}$ is significantly longer than in sample B$5$ where $\tau _D=1.8\ {\rm ns.%
}$ In the right panels of Fig.~\ref{FigData} we show the distribution
functions measured at the lateral position of samples B$10$ (top right
panel) and D$20{\rm b}$ (bottom right panel). The height of the plateau well
agrees with $1-x\approx 0.8,$ and the distribution functions are again more
rounded for the wire with the longer diffusion time.

\begin{figure}[!tb]
\centering
\epsfig{file=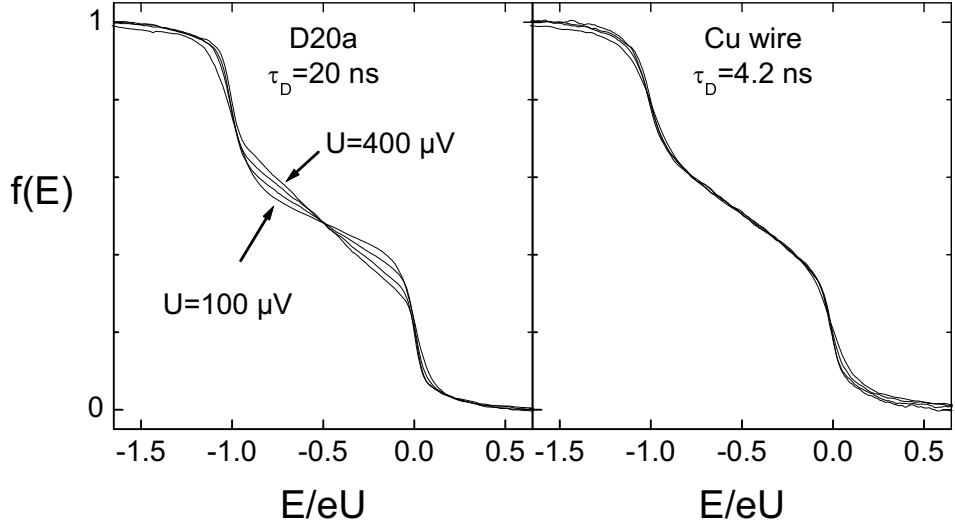, width=4.9692in, clip=}
\caption{
Distribution functions for $U=0.1,$ $0.2,$ $0.3,$ and $0.4\ {\rm mV},$
plotted as a function of the reduced energy $E/eU.$ Left panel: Ag sample D$%
20{\rm a;}$ right panel: Cu sample, $L=5\ {\rm \mu m.}$}
\label{FigScal}
\end{figure}

The distribution functions obtained in the experiments on copper and gold
wires systematically display a scaling property\cite{Relax_1}: as shown for
a copper sample \cite{CuSample} in the right panel of Fig.~\ref{FigScal}, $%
f(x,E)$ only depends, at each position, on the reduced variable $E/eU.$ Such
a scaling law is not observed in our silver samples: in particular, the
slope of the plateau in the distribution functions of the wire D$20{\rm a}$
increases with $U$ when plotted in reduced units (see the left panel of Fig.~%
\ref{FigScal}). This indicates that interactions between quasiparticles have
a different energy dependence in the two types of samples. Interactions are
also weaker in silver samples: as illustrated in Fig.~\ref{FigScal}, similar
distribution functions are obtained for longer diffusion time in silver than
in copper or gold samples.

In order to compare the energy distribution functions $f\left( x,E\right) $
obtained experimentally with the theoretical predictions, we now explain how
the energy exchange rate between quasiparticles determines $f\left(
x,E\right) .$ The stationary distribution function $f\left( x,E\right) $
obeys the Boltzmann equation \cite{Nagaev,Kozub}: 
\begin{equation}
\frac 1{\tau _D}\frac{\partial ^2f\left( x,E\right) }{\partial x^2}+{\cal I}%
_{{\rm coll}}\left( x,E,\left\{ f\right\} \right) =0  \label{Boltzmann}
\end{equation}
where ${\cal I}_{{\rm coll}}\left( x,E,\left\{ f\right\} \right) $ is the
collision integral due to interactions between quasiparticles. The boundary
conditions are imposed by the reservoirs at both ends: $f\left( 0,E\right)
=\left( 1+\exp \frac E{k_BT}\right) ^{-1}$ and $f\left( 1,E\right) =\left(
1+\exp \frac{E+eU}{k_BT}\right) ^{-1}.$

In the absence of inelastic scattering ${\cal I}_{{\rm coll}}=0$ and the
distribution function $f_0\left( x,E\right) $ is \cite{Nagaev}: 
\begin{equation}
f_0\left( x,E\right) =(1-x)f\left( 0,E\right) +xf\left( 1,E\right) .
\label{fzero}
\end{equation}
The function $f_0\left( x,E\right) $ has a well-defined plateau for $\left|
eU\right| \gg k_BT,$ as observed in sample B$5$ (see Fig. 2).

\begin{figure}[!tb]
\centering
\epsfig{file=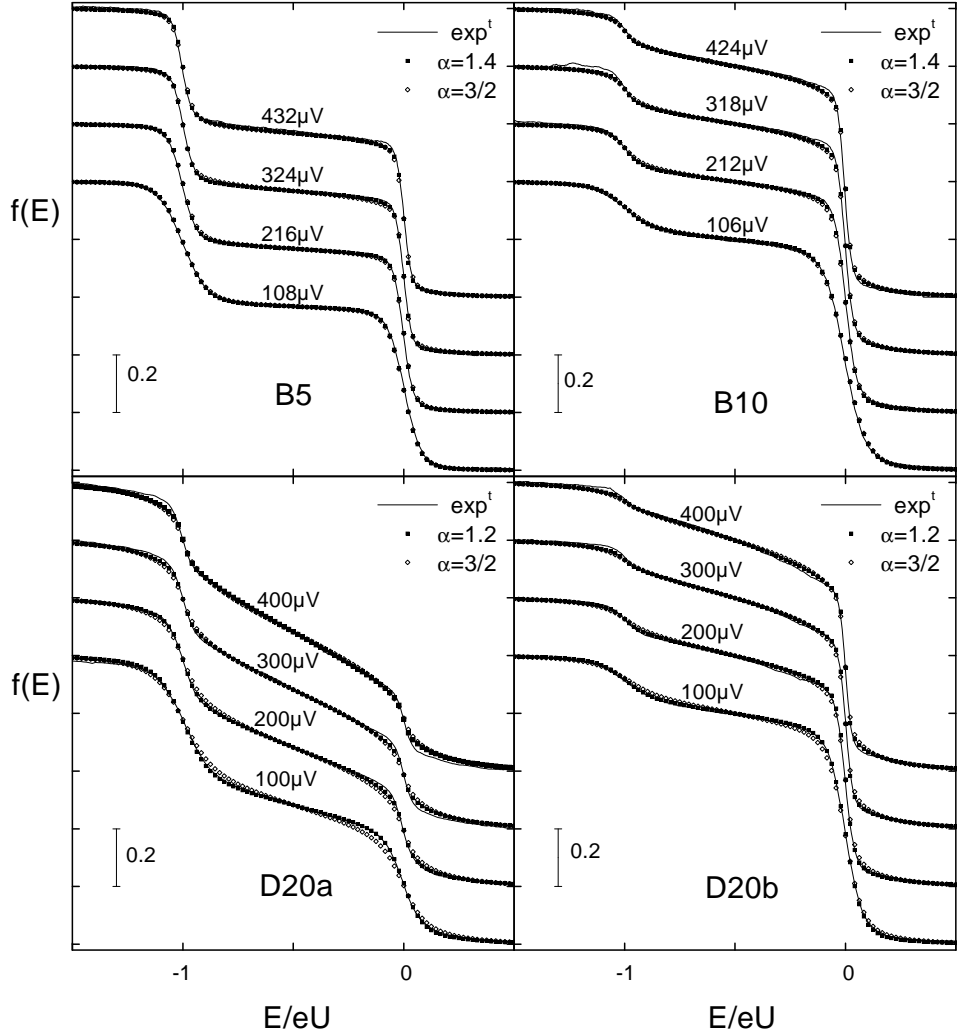, width=4.9787in, clip=}
\caption{Continuous lines in all four
panels: measured distribution functions, plotted as a function of the
reduced variable $E/eU$ for $U=0.1,$ $0.2,$ $0.3$ and $0.4~{\rm mV},$ for
the same samples as in Fig. 2. Successive curves were shifted vertically by
0.2, for clarity. Full squares are the best fits of the data to the solution
of the Boltzmann equation with an interaction kernel $K\left( \varepsilon
\right) =$ $\kappa _{\alpha }\varepsilon ^{-\alpha }.$ Open diamonds are the
best fits obtained with $\alpha $ set to the theoretical value $3/2,$ and
adjusting the prefactor $\kappa _{3/2}.$ Detailed fit parameters are given
in Table~1.}
\label{FigFit}
\end{figure}

The collision term ${\cal I}_{{\rm coll}}\left( x,E,\left\{ f\right\}
\right) $ is the difference of two terms: an in-collision term, the rate at
which particles are scattered into a state of energy $E,$ and an
out-collision term: 
\begin{equation}
{\cal I}_{{\rm coll}}\left( x,E,\left\{ f\right\} \right) ={\cal I}_{{\rm %
coll}}^{{\rm in}}\left( x,E,\left\{ f\right\} \right) -{\cal I}_{{\rm coll}%
}^{{\rm out}}\left( x,E,\left\{ f\right\} \right)   \label{Icoll}
\end{equation}
with 
\begin{eqnarray}
{\cal I}_{{\rm coll}}^{{\rm in,out}}\left( x,E,\left\{ f\right\} \right) 
&=&\int {\rm d}\varepsilon {\rm d}E^{\prime }K\left( \varepsilon \right) 
\label{Iout} \\
&&\times f_{E+\varepsilon ,E}^{x}(1-f_{E,E-\varepsilon }^{x})f_{E^{\prime
}}^{x}(1-f_{E^{\prime }+\varepsilon }^{x})  \nonumber
\end{eqnarray}
where the shorthand $f_{E}^{x}$ stands for $f\left( x,E\right) .$ Following
Landau's approach \cite{Pines}, we have first assumed that the dominant
process is a two-quasiparticle interaction. Moreover the interaction is
assumed to be local on the scale of variations of the distribution function.
The kernel function $K\left( \varepsilon \right) $ is proportional to the
averaged squared interaction between two quasiparticles exchanging an energy 
$\varepsilon .$ The scaling property observed for copper samples implies\cite
{Relax_1} $K\left( \varepsilon \right) \propto \varepsilon ^{-2}$ , as
opposed to the AA prediction $K\left( \varepsilon \right) \propto
\varepsilon ^{-3/2}$ for a diffusive conductor in the $1{\rm D}$ regime\cite
{Altshuler}. In silver samples we have assumed that the interaction kernel
still obeys a power law $K\left( \varepsilon \right) =\kappa _{\alpha
}\varepsilon ^{-\alpha },$ with $\kappa _{\alpha }$ and $\alpha $ taken as
fitting parameters. These best fit theory curves, obtained with the
parameters given in the colums ``fit $\kappa _{\alpha }\varepsilon ^{-\alpha
}$'' of Table~1, are plotted in Fig. \ref{FigFit} with full squares. For
comparison, the best fits obtained with the exponent set at its predicted
value $\alpha =3/2$ are plotted with open diamonds. Note that the same
fitting parameters were used for all the samples of a given experiment${\rm .%
}$\ We obtained the exponent values $\alpha =1.6\pm 0.2$ and $1.4\pm 0.2$
for experiments A and B, respectively, which are compatible with the
prediction $\alpha =3/2$ (the diamonds in the top panels of Fig.~\ref{FigFit}
are practically superimposed with the squares). This compatibility is not
found for the experiments C$\ $and D, for which $\alpha =1.2\pm 0.15:$ the
quality of the fits is visibly degraded by imposing $\alpha =3/2,$ as shown
in the bottom panels of Fig. \ref{FigFit}. The slope of the intermediate
plateau in the theoretical curve (open diamonds) is systematically too large
at $U=100~{\rm \mu V}$ and $U=200~{\rm \mu V}$.%
\begin{table}[tbp] \begin{center}%
\begin{tabular}{|c||c|c||c|c|c||c|p{0.5in}||c|}
\hline
sample & \multicolumn{2}{|c||}{parameters} & \multicolumn{3}{|c||}{fit $%
\kappa _{\alpha }\varepsilon ^{-\alpha }$} & \multicolumn{2}{|c||}{fit TLS+AA
} & theory \\ \hline
& $w$ & $D$ & $\alpha $ & $\kappa _{\alpha }$ & $\kappa _{3/2}$ & $\kappa _{%
{\rm TLS}}$ & $\hspace{0.3cm} \kappa _{3/2}$ & $\kappa _{3/2}^{{\rm thy}}$ \\ \hline
A5 & $90$ & $0.011$ & $1.6_{^{\pm 0.2}}$ & $1.02$ & $1.30$ & $0.7_{^{\pm
0.7}}$ & $1.15_{^{\mp 0.15}}$ & $0.14$ \\ \hline
B5,10 & $65$ & $0.015$ & $1.4_{^{\pm 0.2}}$ & $0.73$ & $0.62$ & $0.9_{^{\pm
0.9}}$ & $0.41_{^{\mp 0.21}}$ & $0.17$ \\ \hline
C20 & $160$ & $0.023$ & $1.2_{^{\pm 0.15}}$ & $1.23$ & $0.70$ & $2.1_{^{\pm
0.5}}$ & $0.19_{^{\mp 0.11}}$ & $0.06$ \\ \hline
D20a,b & $100$ & $0.020$ & $1.2_{^{\pm 0.15}}$ & $0.95$ & $0.49$ & $%
1.5_{^{\pm 0.4}}$ & $0.20_{^{\mp 0.08}}$ & $0.10$ \\ \hline
\end{tabular}
\end{center}
Table 1. Samples parameters and fit parameters. The wire width $w$ is in nm, the diffusion constant $D$ in m$^2\QTR{rm}{s}^{-1}.$ The fit parameters $\kappa _\alpha $ are given in $\QTR{rm}{ns}^{-1}\QTR{rm}{meV}^{\alpha -2},$ and $\kappa _{\QTR{rm}{TLS}}$ is given in $\QTR{rm}{ns}^{-1}\QTR{rm}{meV}^{-1}$.
\end{table}%

In order to explain the discrepancy between the predicted exponent $\alpha
=3/2$ and the value $\alpha =1.2\pm 0.15$ extracted from experiments C and
D, we have investigated theoretically the effect of two-level systems (TLS)
on the energy exchange between quasiparticles. The relevance of TLS on phase
relaxation has recently been suggested by several authors \cite{Imry,Zawa}.
We assume here that the quasiparticles are weakly coupled to the TLS, which
are equally distributed along the wire. These TLS could be atoms moving
across crystalline defects, or impurities in the crystal, for example. The
TLS are assumed to have a flat energy distribution, and to be all equally
coupled to the quasiparticles. We treat the absorption and emission of
energy by the two-level systems in the perturbative limit: the absorption
rate $\Gamma _{+}^x(\varepsilon )$ (respectively the emission rate $\Gamma
_{-}^x(\varepsilon )$) by a TLS at position $x$ with an energy separation $%
\varepsilon $ is given by Fermi's golden rule: $\Gamma _{+}^x(\varepsilon
)=\lambda p_{-}^x(\varepsilon )h^x(\varepsilon )$ (resp. $\Gamma
_{-}^x(\varepsilon )=\lambda p_{+}^x(\varepsilon )h^x(-\varepsilon )$). In
these expressions, $\lambda $ is the coupling constant between
quasiparticles and the TLS, $p_{-}^x(\varepsilon )$ (resp. $%
p_{+}^x(\varepsilon )$) is the occupation probability of the low-energy
level (resp. the high-energy level)$,$ and $h^x(\varepsilon )=\int {\rm d}%
E~f_E^x(1-f_{E-\varepsilon }^x).$ Assuming that the TLS reach a local
equilibrium with the quasiparticles, {\em i.e. }$\Gamma _{+}^x(\varepsilon
)=\Gamma _{-}^x(\varepsilon )$, one obtains $p_{\pm }^x(\varepsilon
)=h^x(\pm \varepsilon )/(h^x(-\varepsilon )+h^x(\varepsilon )).$ Each term $%
\lambda p_{-}(\varepsilon )f_E(1-f_{E-\varepsilon })$ in $\Gamma
_{+}(\varepsilon )$ corresponds to an energy transfer of $\varepsilon $ from
a quasiparticle at energy $E$ to a TLS. Therefore, it gives an ``out''
collision term in the Boltzmann equation (Eq.~(\ref{Boltzmann})) for
quasiparticles at energy $E.$ One then finds directly that the coupling to
TLS can be written as an effective kernel function in Eq. (\ref{Iout}),
which depends on the local distribution function: $K_{{\rm eff}%
}^x(\varepsilon )=\kappa _{{\rm TLS}}/(h^x(-\varepsilon )+h^x(\varepsilon
)). $ The parameter $\kappa _{{\rm TLS}}$ is proportional to the density of
TLS and to the coupling constant $\lambda .$ The effective kernel $K_{{\rm %
eff}}^x(\varepsilon )$ is not a power-law function of $\varepsilon $: at
energies large compared to $eU,$ $K_{{\rm eff}}^x(\varepsilon )\propto
\varepsilon ^{-1},$ but $K_{{\rm eff}}^x(0)$ remains finite. Without
assuming an unrealistic heating of the reservoirs, it was not possible to
fit the data with this model alone. However, we found that all the
experimental data could be well accounted for by assuming the presence of
two phenomena: direct quasiparticle-quasiparticle interaction, described by
the AA theory \cite{Altshuler}: $K(\varepsilon )=\kappa _{3/2}\varepsilon
^{-3/2};$ and quasiparticle-TLS coupling, described by $K_{{\rm eff}%
}^x(\varepsilon ).$ The parameters $\kappa _{{\rm TLS}}$ and $\kappa _{3/2}$
for the best fits (which are not shown because they are hardly
distinguishable from the fits with the best value of $\alpha $ in Fig.~4)
are given in the columns ``fits TLS+AA'' of Table~1. Note that the error
bars on $\kappa _{{\rm TLS}}$ and $\kappa _{3/2}$ are correlated: the
weighted sum of $\kappa _{{\rm TLS}}$ and $\kappa _{3/2}$ must remain
constant. Apart from sample A5, the theoretical value for $\kappa _{3/2}^{%
{\rm thy}}$ are of the same order of magnitude as $\kappa _{3/2}.$ For the
experiments A and B, extra contributions to AA theory are minimal. In
contrast, in the experiments C and D the slope of the plateau in the
distribution function is well explained by the TLS, whereas the AA
mechanism, which dominates at low energy, is responsible for the rounding of
the steps.

The differences that we observed between copper, gold and silver wires in
the energy exchange rate experiments were also found in measurements of the
phase coherence time $\tau _\phi $ in wires fabricated using the same
procedure. In a silver sample, the temperature dependence of $\tau _\phi $
follows closely the theoretical prediction down to the base temperature $%
T=50 $ ${\rm mK}$ of the refrigerator (where $\tau _\phi =9\ {\rm ns}$),
whereas $\tau _\phi $ saturates below $1$ ${\rm K}$ at $\tau _\phi =1~%
{\rm ns}$ in our copper sample, and below $6$ ${\rm K}$ at $\tau _\phi
=10~{\rm ps}$ in our gold sample. These measurements are further discussed
in this volume\cite{Birge}~.

In conclusion, we have found that interactions between quasiparticles in
silver wires are much weaker and have a different qualitative behavior than
in copper or gold wires. The energy exchange rate in our silver samples is
close to the theoretical predictions for a diffusive medium in the $1{\rm D}$
regime, provided that one includes an extra contribution, which might be due
to two-level systems. The difference in the behavior of interactions in
gold, copper and silver samples, as seen from the energy exchange rate
experiments, is correlated to the measured saturation of $\tau _\phi $ at
low temperature which did occur in copper and gold, but not in silver. A
possible interpretation could be that two-level systems are more numerous
and/or better coupled to quasiparticles in copper and gold than in silver,
leading to faster phase and energy relaxation. Moreover, preliminary
calculations in the strong coupling regime account for the scaling property
of the distribution functions found in copper and gold samples\cite{Zawa}.
Experiments are in progress to test this interpretation.

Acknowledgments: We are grateful to Norman Birge, P.\ Joyez, H.\ Kroha and
C.\ Urbina for useful discussions and comments, and to P.F.\ Orfila for
technical assistance. We acknowledge A.\ Steinbach for carefully reading the
manuscript. This work has been partly founded by the Bureau National de la
M\'{e}trologie.


\begin{thebibliography}{99}
\bibitem{Treservoir}  The best fit was obtained with a reservoir temperature
of $80~{\rm mK.}$
\bibitem{CuSample}  The width of this $5$-$\mu $m-long copper wire was $w=95~%
{\rm nm,}$ and the diffusion constant $D=0.0065$~{\rm m}$^{2}{\rm s}^{-1}.$
\end{thebibliography}
\end{document}